%Paper: cond-mat/9303035
%From: John Cardy <cardy@sarek.physics.ucsb.edu>
%Date: Fri, 19 Mar 93 13:48:59 -0800

%% UNIVERSAL AMPLITUDE RATIOS FOR SELF-AVOIDING WALKS, POLYGONS & TRAILS %%%

%%%%%%%%%%%%%%%% JOHN L. CARDY & ANTHONY J. GUTTMANN %%%%%%%%%%%%%%%%%%%%%%

%%%%%%%%%%%%%%%% Use Plain TeX. No macros required. %%%%%%%%%%%%%%%%%%%%%%%%

\font\twelverm=cmr10 scaled 1200    \font\twelvei=cmmi10 scaled 1200
\font\twelvesy=cmsy10 scaled 1200   \font\twelveex=cmex10 scaled 1200
\font\twelvebf=cmbx10 scaled 1200   \font\twelvesl=cmsl10 scaled 1200
\font\twelvett=cmtt10 scaled 1200   \font\twelveit=cmti10 scaled 1200

\skewchar\twelvei='177   \skewchar\twelvesy='60
\def\twelvepoint{\normalbaselineskip=12.4pt
  \abovedisplayskip 12.4pt plus 3pt minus 9pt
  \belowdisplayskip 12.4pt plus 3pt minus 9pt
  \abovedisplayshortskip 0pt plus 3pt
  \belowdisplayshortskip 7.2pt plus 3pt minus 4pt
  \smallskipamount=3.6pt plus1.2pt minus1.2pt
  \medskipamount=7.2pt plus2.4pt minus2.4pt
  \bigskipamount=14.4pt plus4.8pt minus4.8pt
  \def\rm{\fam0\twelverm}          \def\it{\fam\itfam\twelveit}%
  \def\sl{\fam\slfam\twelvesl}     \def\bf{\fam\bffam\twelvebf}%
  \def\mit{\fam 1}                 \def\cal{\fam 2}%
  \def\tt{\twelvett}
  \textfont0=\twelverm   \scriptfont0=\tenrm   \scriptscriptfont0=\sevenrm
  \textfont1=\twelvei    \scriptfont1=\teni    \scriptscriptfont1=\seveni
  \textfont2=\twelvesy   \scriptfont2=\tensy   \scriptscriptfont2=\sevensy
  \textfont3=\twelveex   \scriptfont3=\twelveex  \scriptscriptfont3=\twelveex
  \textfont\itfam=\twelveit
  \textfont\slfam=\twelvesl
  \textfont\bffam=\twelvebf \scriptfont\bffam=\tenbf
  \scriptscriptfont\bffam=\sevenbf
  \normalbaselines\rm}

\def\beginlinemode{\endmode
  \begingroup\parskip=0pt \obeylines\def\\{\par}\def\endmode{\par\endgroup}}
\def\beginparmode{\endmode
  \begingroup \def\endmode{\par\endgroup}}
\let\endmode=\par
{\obeylines\gdef\
{}}
\def\singlespace{\baselineskip=\normalbaselineskip}
\def\oneandathirdspace{\baselineskip=\normalbaselineskip
  \multiply\baselineskip by 4 \divide\baselineskip by 3}
\def\oneandahalfspace{\baselineskip=\normalbaselineskip
  \multiply\baselineskip by 3 \divide\baselineskip by 2}
\def\doublespace{\baselineskip=\normalbaselineskip \multiply\baselineskip by 2}

\newcount\firstpageno
\firstpageno=2
%% FOLLOWING LINE CANNOT BE BROKEN BEFORE 80 CHAR
\footline={\ifnum\pageno<\firstpageno{\hfil}\else{\hfil\twelverm\folio\hfil}\fi}
\let\rawfootnote=\footnote		% We must set the footnote style
\def\footnote#1#2{{\rm\singlespace\parindent=0pt\rawfootnote{#1}{#2}}}
\def\raggedcenter{\leftskip=4em plus 12em \rightskip=\leftskip
  \parindent=0pt \parfillskip=0pt \spaceskip=.3333em \xspaceskip=.5em
  \pretolerance=9999 \tolerance=9999
  \hyphenpenalty=9999 \exhyphenpenalty=9999 }
\hsize=6.5truein
\vsize=8.9truein
\parskip=\medskipamount
\twelvepoint		% selects twelvepoint fonts (cf. \tenpoint)
\doublespace		% selects double spacing for main part of paper (cf.
			%	\singlespace, \oneandahalfspace)
\overfullrule=0pt	% delete the nasty little black boxes for overfull box
\def\preprintno#1{
 \rightline{\rm #1}}	% Preprint number at upper right of title page

\def\title			%  Title on title page
  {\null\vskip 3pt plus 0.2fill
   \beginlinemode \doublespace \raggedcenter \bf}

\def\author			%  Author(s) name(s)  on title page
  {\vskip 4pt plus 0.2fill \beginlinemode
   \singlespace \raggedcenter}

\def\affil			% Affiliations (can intermix with \author)
  {\vskip 3pt plus 0.1fill \beginlinemode
   \oneandahalfspace \raggedcenter \sl}

\def\abstract			% Begin abstract
  {\vskip 3pt plus 0.3fill \beginparmode
%  \doublespace \narrower ABSTRACT: }
   \doublespace \narrower }

\def\endtitlepage		% End title page, begin body of paper
  {\endpage			% 	This subsumes \body
   \body}

\def\body			% Begin text body;  can be used to end
  {\beginparmode}		% \title, \author, \affil, \abstract,
				% \reference, or \figurecaption modes

\def\head#1{			% Head;  NOTE enclose the text in {}
  \filbreak\vskip 0.5truein	%  e.g., \head{I. Introduction}
  {\immediate\write16{#1}
   \line{\bf{#1}\hfil}\par}
   \nobreak\vskip 0.25truein\nobreak}

\def\refto#1{$^{#1}$}		% For references in text as superscript

\def\references			% Begin references -- basic format is Phys Rev
  {\head{References}		% I.e., volume, page, year (space after commas).
   \beginparmode
   \frenchspacing \parindent=0pt \leftskip=0truecm
   \parskip=8pt plus 3pt \everypar{\hangindent=\parindent}}

\gdef\refis#1{\indent\hbox to 0pt{\hss#1.~}}	% Ref list numbers.

\gdef\journal#1, #2, #3, 1#4#5#6{		% Journal reference.  Comma sets
    {\sl #1~}{\bf #2}, #3, (1#4#5#6)}		% off: name, vol, page, year

\def\refstylenp{		% Nucl Phys(or Phys Lett) ref style: V, Y, P
  \gdef\refto##1{ [##1]}				% Reference in text []
  \gdef\refis##1{\indent\hbox to 0pt{\hss##1)~}}	% Ref list numbers)
  \gdef\journal##1, ##2, ##3, ##4 {			% Journal reference
     {\sl ##1~}{\bf ##2~}(##3) ##4 }}

\def\endreferences{\body}
\def\endpage			%  Eject a page
  {\vfill\eject}
\def\endpaper                   %  Ways to say goodbye
  {\endmode\vfill\supereject}
\def\endit
  {\endpaper\end}
\def\ref#1{Ref.\thinspace#1}			% 	for inline references

\def\Ref#1{Ref.\thinspace#1}			% 	ditto

			%	ditto
			%	ditto
			%	ditto
			%	ditto

\def\(#1){Eq.\thinspace(\call{#1})}
\def\call#1{{#1}}

\def\frac#1#2{{#1 \over #2}}

\def\ffrac#1#2{\textstyle{#1\over#2}\displaystyle}

\refstylenp

\catcode`@=11
\newcount\r@fcount \r@fcount=0
\newcount\r@fcurr
\immediate\newwrite\reffile
\newif\ifr@ffile\r@ffilefalse
\def\w@rnwrite#1{\ifr@ffile\immediate\write\reffile{#1}\fi\message{#1}}

\def\writer@f#1>>{}
\def\referencefile{%			  Stuff to write .REF file
  \r@ffiletrue\immediate\openout\reffile=\jobname.ref%
  \def\writer@f##1>>{\ifr@ffile\immediate\write\reffile%
    {\noexpand\refis{##1} = \csname r@fnum##1\endcsname = %
     \expandafter\expandafter\expandafter\strip@t\expandafter%
     \meaning\csname r@ftext\csname r@fnum##1\endcsname\endcsname}\fi}%
  \def\strip@t##1>>{}}

\def\citeall#1{\xdef#1##1{#1{\noexpand\cite{##1}}}}
\def\cite#1{\each@rg\citer@nge{#1}}	% Variable No. of args, separated by ","

\def\each@rg#1#2{{\let\thecsname=#1\expandafter\first@rg#2,\end,}}
\def\first@rg#1,{\thecsname{#1}\apply@rg}	% each@ag is a general purpose
\def\apply@rg#1,{\ifx\end#1\let\next=\relax%	  variable no. of arg. macro.
\else,\thecsname{#1}\let\next=\apply@rg\fi\next}% args separated by commas

\def\citer@nge#1{\citedor@nge#1-\end-}	% Check for M-N range (M and N numbers)
\def\citer@ngeat#1\end-{#1}
\def\citedor@nge#1-#2-{\ifx\end#2\r@featspace#1 % Single argument
  \else\citel@@p{#1}{#2}\citer@ngeat\fi}	% M-N range of arguments
\def\citel@@p#1#2{\ifnum#1>#2{\errmessage{Reference range #1-#2\space is bad.}%
    \errhelp{If you cite a series of references by the notation M-N, then M and
    N must be integers, and N must be greater than or equal to M.}}\else%
 {\count0=#1\count1=#2\advance\count1
by1\relax\expandafter\r@fcite\the\count0,%
  \loop\advance\count0 by1\relax%	  Loop from M to N
    \ifnum\count0<\count1,\expandafter\r@fcite\the\count0,%
  \repeat}\fi}

\def\r@featspace#1#2 {\r@fcite#1#2,}	% Eat spaces at beginning or end of arg
\def\r@fcite#1,{\ifuncit@d{#1}%		  Cite individual reference
    \newr@f{#1}%
    \expandafter\gdef\csname r@ftext\number\r@fcount\endcsname%
                     {\message{Reference #1 to be supplied.}%
                      \writer@f#1>>#1 to be supplied.\par}%
 \fi%
 \csname r@fnum#1\endcsname}
\def\ifuncit@d#1{\expandafter\ifx\csname r@fnum#1\endcsname\relax}%
\def\newr@f#1{\global\advance\r@fcount by1%
    \expandafter\xdef\csname r@fnum#1\endcsname{\number\r@fcount}}

\let\r@fis=\refis			% Save old \refis, redefine
\def\refis#1#2#3\par{\ifuncit@d{#1}%      Use two params #2 #3 to strip blank
   \newr@f{#1}%
   \w@rnwrite{Reference #1=\number\r@fcount\space is not cited up to now.}\fi%
  \expandafter\gdef\csname r@ftext\csname r@fnum#1\endcsname\endcsname%
  {\writer@f#1>>#2#3\par}}

\def\ignoreuncited{%   redefine \refis if ignoring uncited references
   \def\refis##1##2##3\par{\ifuncit@d{##1}%
     \else\expandafter\gdef\csname r@ftext\csname
r@fnum##1\endcsname\endcsname%
     {\writer@f##1>>##2##3\par}\fi}}

\def\r@ferr{\endreferences\errmessage{I was expecting to see
\noexpand\endreferences before now;  I have inserted it here.}}
\let\r@ferences=\references
\def\references{\r@ferences\def\endmode{\r@ferr\par\endgroup}}

\let\endr@ferences=\endreferences
\def\endreferences{\r@fcurr=0%		  Save old \endreferences, redefine
  {\loop\ifnum\r@fcurr<\r@fcount%	  Loop over refnum and produce text
    \advance\r@fcurr by 1\relax\expandafter\r@fis\expandafter{\number\r@fcurr}%
    \csname r@ftext\number\r@fcurr\endcsname%
  \repeat}\gdef\r@ferr{}\endr@ferences}

% Save old \endpaper, redefine it to write parting message.

\let\r@fend=\endpaper\gdef\endpaper{\ifr@ffile
\immediate\write16{Cross References written on []\jobname.REF.}\fi\r@fend}

\catcode`@=12

\citeall\refto		% These macros will generate citations
\citeall\ref		%
\citeall\Ref		%

\catcode`@=11
\newcount\tagnumber\tagnumber=0

\immediate\newwrite\eqnfile
\newif\if@qnfile\@qnfilefalse
\def\write@qn#1{}
\def\writenew@qn#1{}
\def\w@rnwrite#1{\write@qn{#1}\message{#1}}
\def\@rrwrite#1{\write@qn{#1}\errmessage{#1}}

\def\taghead#1{\gdef\t@ghead{#1}\global\tagnumber=0}
\def\t@ghead{}

\expandafter\def\csname @qnnum-3\endcsname
  {{\t@ghead\advance\tagnumber by -3\relax\number\tagnumber}}
\expandafter\def\csname @qnnum-2\endcsname
  {{\t@ghead\advance\tagnumber by -2\relax\number\tagnumber}}
\expandafter\def\csname @qnnum-1\endcsname
  {{\t@ghead\advance\tagnumber by -1\relax\number\tagnumber}}
\expandafter\def\csname @qnnum0\endcsname
  {\t@ghead\number\tagnumber}
\expandafter\def\csname @qnnum+1\endcsname
  {{\t@ghead\advance\tagnumber by 1\relax\number\tagnumber}}
\expandafter\def\csname @qnnum+2\endcsname
  {{\t@ghead\advance\tagnumber by 2\relax\number\tagnumber}}
\expandafter\def\csname @qnnum+3\endcsname
  {{\t@ghead\advance\tagnumber by 3\relax\number\tagnumber}}

\def\equationfile{%
  \@qnfiletrue\immediate\openout\eqnfile=\jobname.eqn%
  \def\write@qn##1{\if@qnfile\immediate\write\eqnfile{##1}\fi}
  \def\writenew@qn##1{\if@qnfile\immediate\write\eqnfile
    {\noexpand\tag{##1} = (\t@ghead\number\tagnumber)}\fi}
}

\def\callall#1{\xdef#1##1{#1{\noexpand\call{##1}}}}
\def\call#1{\each@rg\callr@nge{#1}}

\def\each@rg#1#2{{\let\thecsname=#1\expandafter\first@rg#2,\end,}}
\def\first@rg#1,{\thecsname{#1}\apply@rg}
\def\apply@rg#1,{\ifx\end#1\let\next=\relax%
\else,\thecsname{#1}\let\next=\apply@rg\fi\next}

\def\callr@nge#1{\calldor@nge#1-\end-}
\def\callr@ngeat#1\end-{#1}
\def\calldor@nge#1-#2-{\ifx\end#2\@qneatspace#1 %
  \else\calll@@p{#1}{#2}\callr@ngeat\fi}
\def\calll@@p#1#2{\ifnum#1>#2{\@rrwrite{Equation range #1-#2\space is bad.}
\errhelp{If you call a series of equations by the notation M-N, then M and
N must be integers, and N must be greater than or equal to M.}}\else%
 {\count0=#1\count1=#2\advance\count1
by1\relax\expandafter\@qncall\the\count0,%
  \loop\advance\count0 by1\relax%
    \ifnum\count0<\count1,\expandafter\@qncall\the\count0,%
  \repeat}\fi}

\def\@qneatspace#1#2 {\@qncall#1#2,}
\def\@qncall#1,{\ifunc@lled{#1}{\def\next{#1}\ifx\next\empty\else
  \w@rnwrite{Equation number \noexpand\(>>#1<<) has not been defined yet.}
  >>#1<<\fi}\else\csname @qnnum#1\endcsname\fi}

\let\eqnono=\eqno
\def\eqno(#1){\tag#1}
\def\tag#1$${\eqnono(\displayt@g#1 )$$}

\def\aligntag#1$${\gdef\tag##1\\{&(\displayt@g##1 )\cr}\eqalignno{#1\\}$$
  \gdef\tag##1$${\eqnono(\displayt@g##1 )$$}}

\def\eqalignno#1{\displ@y \tabskip\centering
  \halign to\displaywidth{\hfil$\displaystyle{##}$\tabskip\z@skip
    &$\displaystyle{{}##}$\hfil\tabskip\centering
    &\llap{$\displayt@gpar##$}\tabskip\z@skip\crcr
    #1\crcr}}

\def\displayt@gpar(#1){(\displayt@g#1 )}

\def\displayt@g#1 {\rm\ifunc@lled{#1}\global\advance\tagnumber by1
        {\def\next{#1}\ifx\next\empty\else\expandafter
        \xdef\csname @qnnum#1\endcsname{\t@ghead\number\tagnumber}\fi}%
  \writenew@qn{#1}\t@ghead\number\tagnumber\else
        {\edef\next{\t@ghead\number\tagnumber}%
        \expandafter\ifx\csname @qnnum#1\endcsname\next\else
        \w@rnwrite{Equation \noexpand\tag{#1} is a duplicate number.}\fi}%
  \csname @qnnum#1\endcsname\fi}

\def\ifunc@lled#1{\expandafter\ifx\csname @qnnum#1\endcsname\relax}

\let\@qnend=\end\gdef\end{\if@qnfile
\immediate\write16{Equation numbers written on []\jobname.EQN.}\fi\@qnend}

\catcode`@=12

% +--------------------------------------------------------------------+
% |                                                                    |
% |                           TABLES.TEX                               |
% |                                                                    |
% |                     Ray F. Cowan  15-Feb-85                        |
% |                                                                    |
% |                       Princeton University                         |
% |                                                                    |
% |                     Last Revision: 14-Mar-85                       |
% |                                                                    |
% |   Macros I find handy for making tables.  See TABLEDOC TEX for     |
% |   a longer description.  The token-counting macros are straight    |
% |   from the TeXbook's "Dirty Tricks" appendix.                      |
% |                                                                    |
% +--------------------------------------------------------------------+
%
\newbox\hdbox%
\newcount\hdrows%
\newcount\multispancount%
\newcount\ncase%
\newcount\ncols% This is the number of primary text columns in the table.
\newcount\nrows%
\newcount\nspan%
\newcount\ntemp%
\newdimen\hdsize%
\newdimen\newhdsize%
\newdimen\parasize%
\newdimen\thicksize%
\newdimen\thinsize%
\newif\ifendsize%
\newif\iffirstrow%
\newtoks\dbt%
\newtoks\hdtks%
\newtoks\savetks%
\newtoks\tableLETtokens%
\newtoks\tabletokens%
%
%  Book-keeping stuff--see how often these macros are called.
%
\immediate\write15{%
CP SMSG GJMSINK TEXTABLE -----> TABLE MACROS LOADED, JOB = \jobname%
}%
\catcode`\@=11%  Allows use of "@" in macro names, like PLAIN TEX does.
%  Debugging aid.  Writes #1 on the
%                                    user's terminal and in the log file.
\def\tstrut{\vrule height16pt depth6pt width0pt}%
\def\|{|}%  Make it easy to get |'s of type other after they are later
%           made active.
\def\tablerule{\noalign{\hrule height\thinsize depth0pt}}%
\thicksize=1.5pt%  Default thickness for fat rules.  The user should feel
%                  free to change this to his preference.
\thinsize=0.6pt%   Default thickness for thin rules.
\def\thickrule{\noalign{\hrule height\thicksize depth0pt}}%
\def\ctr#1{\hfil\ #1\ \hfil}%
%
%  \vctr vertically centers its argument in the row.
\def\tablewidth{}%
\parasize=4in%
\gdef\ARGS{########}%  Produces the correct number of #'s in the preamble
%                      by the time eveything is expanded and \halign sees
%                      it.
\gdef\headerARGS{####}%  Same as \ARGS, but used in \header macros.
\def\@mpersand{&}%  Allows us to get alignment tab characters later
%                   when we have made the character "&" an active macro.
{\catcode`\|=13%  Make |'s locally active.
\gdef\letbarzero{\let|0}%  Globally define a macro that allows us to
%                          keep active |'s from being expanded in edef's.
\gdef\letbartab{\def|{&&}}%  This \def will cause active |'s read by
%                            \ruledtable to be converted into double
%                            alignment tabs.
}%  End of locally active |'s.
{\def\ampskip{&\omit\hfil&}%  This local macro skips a vertical rule.
\catcode`\&=13%  Now make &'s into active macros.
\let&0%  This allows us to expand \ampskip in the next \xdef without
%        attempting to expand the & and getting an "undefined control
%        sequence" error.
\xdef\letampskip{\def&{\ampskip}}%  This will cause active &'s read by
%                                   \ruledtable to be converted into
%                                   double tabs and an \omit'ted \vrule.
}%  End of locally active &'s.
\def\begintable{%  Here we make |'s and &'s active characters so we can
%                  interpret them as macros.  Note that this action is
%                  true only until we encounter the matching \endgroup
%                  token later at the end of the \ruledtable macro.
   \begingroup%
   \catcode`\|=13\letbartab%
   \catcode`\&=13\letampskip%
   \def\multispan##1{%  We must redefine \multispan to count the number
%                       of primary columns, not physical columns.
      \omit \mscount##1%
      \multiply\mscount\tw@\advance\mscount\m@ne%
      \loop\ifnum\mscount>\@ne \sp@n\repeat%
   }%  End of \multispan macro.
   \def\|{%
      &\omit\widevline&%
   }%
   \ruledtable%  Now we call \ruledtable to do the real work.
}%  End of \begintable macro.
\newif\iflatetable \latetablefalse

\newbox\tablebox
% lets see if this works
\long\def\ruledtable#1\endtable{%
%
%  This macro reads in the user's data entries
%  and converts them into a ruled table.
%
%  Important note:  Many macros and parameters are re-defined here, and
%  these must be kept local to the table macros to avoid conflict with
%  their use outside of tables.  This is done by the \begingroup token
%  macro \begintable and the \endgroup token at the end of
%  this macro.
%
   \offinterlineskip%  Needed to make rules touch each other.
   \tabskip 0pt%  Needed for same reason as \offinterlineskip.
   \def\widevline{\vrule width\thicksize}%  Make outer \vrule's wider.
   \def\endrow{\@mpersand\omit\hfil\crnorm\@mpersand}%
   \def\crthick{\@mpersand\crnorm\thickrule\@mpersand}%
   \def\crnorule{\@mpersand\crnorm\@mpersand}%
   \let\nr=\crnorule%  A shorter abbreviation.
   \def\endtable{\@mpersand\crnorm\thickrule}%
   \let\crnorm=\cr%  Allows us to use \cr for our own purposes.
%
%  Cause user-typed \cr's to follow a row with a \tablerule.
%
   \edef\cr{\@mpersand\crnorm\tablerule\@mpersand}%
   \the\tableLETtokens%  Get the user's extra \let's, if any.
%
%  Put the data entries into a token register so we can scan through them
%  and see what the user is asking us to do.
%
   \tabletokens={&#1}%  We add an extra alignment tab to the beginning
%                       of the first row to allow for the first \vrule.
%
%  Now count how many rows are in the table and return the result in
%  count register \nrows; do the same for columns, and return that
%  in register \ncols.
%
   \countROWS\tabletokens\into\nrows%
   \countCOLS\tabletokens\into\ncols%
%
%  Now do a little arithmetic to convert the number of primary columns
%  into the number of physical columns that the alignment preamble must
%  prepare for;  similarly for rows.
%
   \advance\ncols by -1%
   \divide\ncols by 2%
   \advance\nrows by 1%
%
%  Tell the user how many rows and columns we found in his data.
%
   \immediate\write16{[Nrows=\the\nrows, Ncols=\the\ncols]}%
%
%  Now we actually go ahead and produce the table.
%
\iflatetable
   \global\setbox\tablebox=\hbox{%
	\tableframe%
	}% End of \hbox
\else
   \tableframe \fi
   \endgroup%  Return all local macros and parameters to their outside
%              values.
}%  End of macro \ruledtable.
\def\tableframe{%
   \line{%  The final table comes out as an \hbox of width the \hsize.
      \hss%  The final table will be centered left-to-right.
	\vbox{%
         \makePREAMBLE{\the\ncols}%  Generate the preamble.
         \edef\next{\preamble}%  This line and the next line force the
         \let\preamble=\next%    expansion of all \ARGS tokens into the
%                                appropriate number of #'s.
         \makeTABLE{\preamble}{\tabletokens}%  Go do the \halign here.
      }%  End of \vbox.
      \hss%  Finish the centering effect.
   }%  End of \line.
	}%	End of tableframe
\def\makeTABLE#1#2{%  Does an \halign for the \ruledtable macro.
   {%  Start of local parameter values.
   \let\ifmath0%     These macros would cause trouble if they were to be
   \let\header0%     expanded in the following \xdef; we \let them be
   \let\multispan0%  equal to a digit, because digits can't be expanded.
   \xdef\next{%  We must force the preamble to be expanded BEFORE the
      \halign\tablewidth{%
%        \halign is done;  this \edef\next{...}\next construction
%                does the trick.
      #1%  This is the preamble text.
      \noalign{\hrule height\thicksize depth0pt}%  Makes the top \hrule.
      \the#2\endtable%  This is the main body.
%
%     \noalign{\hrule height0.7pt depth0pt}%  Makes the last \hrule.
      }%  End of \halign.
   }%  End of \next.
   }%  End of local values.
   \next%  This \next must be outside of the local values, because now
%          we want those troublesome macros in the \let's above to have
%          their normal actions.
}%  End of macro \makeTABLE.
\def\makePREAMBLE#1{%  This macro generates the necessary preamble for a
%                      ruled table with #1 primary columns.
%                      (Primary columns means the number of columns NOT
%                       counting those used for vertical rules.)
   \ncols=#1%  Get the number of columns desired.
   \begingroup%  Start local parameter definitions.
   \let\ARGS=0%  This is the key to the whole thing; it prevents \ARGS
%                from being expanded in the followin \edef's.
   \edef\xtp{\widevline\ARGS&\tstrut\ctr{\ARGS}}%  A 1-column preamble.
   \advance\ncols by -1%  One column has been generated; decrement the
%                         counter.
   \loop%  Append as many further columns as needed to the preamble.
      \ifnum\ncols>0 %
      \advance\ncols by -1%
      \edef\xtp{\xtp&\vrule width\thinsize\ARGS&\ctr{\ARGS}}%
   \repeat
   \xdef\preamble{\xtp&\widevline\ARGS\crnorm}%  Adds the last \vrule.
   \endgroup%  End of local parameters.
}%  End of macro \makePREAMBLE.
\def\countROWS#1\into#2{%  This counts the number of rows in #1 by
%                          looking for control sequences that end a row,
%                          e.g., \cr, \crthick, etc., and puts the result
%                          into count register #2.
   \let\countREGISTER=#2%
   \countREGISTER=0%
%  \out{In countROWS:  tokens are [\the#1]}%
   \expandafter\ROWcount\the#1\endcount%
}%
\def\ROWcount{%
   \afterassignment\subROWcount\let\next= %
}%
\def\subROWcount{%
%  \out{In subROWcount:  next is [\meaning\next]}%  Debugging aid.
   \ifx\next\endcount %
      \let\next=\relax%
   \else%
      \ncase=0%
      \ifx\next\cr %
         \global\advance\countREGISTER by 1%
         \ncase=0%
      \fi%
      \ifx\next\endrow %
         \global\advance\countREGISTER by 1%
         \ncase=0%
      \fi%
      \ifx\next\crthick %
         \global\advance\countREGISTER by 1%
         \ncase=0%
      \fi%
      \ifx\next\crnorule %
         \global\advance\countREGISTER by 1%
         \ncase=0%
      \fi%
      \ifx\next\header %
%     \out{In subROWcount:  next=header, ncase set=1}%
         \ncase=1%
      \fi%
%     \out{In subROWcount:  ncase is [\the\ncase]}%
      \relax%
      \ifcase\ncase %
         \let\next\ROWcount%
%        \out{subROWcount---> ncase=\the\ncase}%
      \or %
         \let\next\argROWskip%
%        \out{subROWcount---> ncase=\the\ncase}%
      \else %
      \fi%
   \fi%
%  \out{subROWcount---> NEXT=\meaning\next}%
   \next%
}%  End of macro \subROWcount.
\def\counthdROWS#1\into#2{%
\dvr{10}%
   \let\countREGISTER=#2%
   \countREGISTER=0%
\dvr{11}%
%  \out{In counthdROWS:  tokens are [\the#1]}%
\dvr{13}%
   \expandafter\hdROWcount\the#1\endcount%
\dvr{12}%
}%
\def\hdROWcount{%
   \afterassignment\subhdROWcount\let\next= %
}%
\def\subhdROWcount{%
%\out{In subhdROWcount:  next is [\meaning\next]}%
   \ifx\next\endcount %
      \let\next=\relax%
   \else%
      \ncase=0%
      \ifx\next\cr %
         \global\advance\countREGISTER by 1%
         \ncase=0%
      \fi%
      \ifx\next\endrow %
         \global\advance\countREGISTER by 1%
         \ncase=0%
      \fi%
      \ifx\next\crthick %
         \global\advance\countREGISTER by 1%
         \ncase=0%
      \fi%
      \ifx\next\crnorule %
         \global\advance\countREGISTER by 1%
         \ncase=0%
      \fi%
      \ifx\next\header %
%\out{In subhdROWcount:  next=header, ncase set=1}%
         \ncase=1%
      \fi%
%\out{In subhdROWcount:  ncase is [\the\ncase]}%
\relax%
      \ifcase\ncase %
         \let\next\hdROWcount%
%\out{subhdROWcount---> ncase=\the\ncase}%
      \or%
         \let\next\arghdROWskip%
%\out{subhdROWcount---> ncase=\the\ncase}%
      \else %
      \fi%
   \fi%
%\out{subhdROWcount---> NEXT=\meaning\next}%
   \next%
}%
{\catcode`\|=13\letbartab
\gdef\countCOLS#1\into#2{%
%  \out{In countCOLS:  tokens are [\the#1]}
   \let\countREGISTER=#2%
   \global\countREGISTER=0%
   \global\multispancount=0%
   \global\firstrowtrue
   \expandafter\COLcount\the#1\endcount%
   \global\advance\countREGISTER by 3%
   \global\advance\countREGISTER by -\multispancount
%  \out{countCOLS-->[\the\countREGISTER]}
}%
\gdef\COLcount{%
   \afterassignment\subCOLcount\let\next= %
}%
{\catcode`\&=13%
\gdef\subCOLcount{%
%\out{In subCOLcount: next is [\meaning\next]}
   \ifx\next\endcount %
      \let\next=\relax%
   \else%
      \ncase=0
      \iffirstrow
         \ifx\next& %
            \global\advance\countREGISTER by 2%
            \ncase=0%
         \fi%
         \ifx\next\span %
            \global\advance\countREGISTER by 1%
            \ncase=0%
         \fi%
         \ifx\next| %
            \global\advance\countREGISTER by 2%
            \ncase=0%
         \fi
         \ifx\next\|
            \global\advance\countREGISTER by 2
            \ncase=0%
         \fi
         \ifx\next\multispan
            \ncase=1%
            \global\advance\multispancount by 1%
         \fi
         \ifx\next\header
            \ncase=2%
         \fi
         \ifx\next\cr       \global\firstrowfalse \fi
         \ifx\next\endrow   \global\firstrowfalse \fi
         \ifx\next\crthick  \global\firstrowfalse \fi
         \ifx\next\crnorule \global\firstrowfalse \fi
      \fi%  End of \iffirstrow.
\relax%\out{subCOL-->  ncase=[\the\ncase]}
% \out{subCOL-->  next=\meaning\next}
      \ifcase\ncase %
         \let\next\COLcount%
      \or %
         \let\next\spancount%
      \or %
         \let\next\argCOLskip%
      \else %
      \fi %
   \fi%
%  \out{subCOL-->  countREGISTER=[\the\countREGISTER]}
   \next%
}%
\gdef\argROWskip#1{%
%  Deletes the next balanced, undelimited argument from a
%                 token list.
% \out{---> Entering argROWskip <---}
% \out{In argROWskip:  deleted arg is [#1]}%
   \let\next\ROWcount \next%
}%  End of macro \argskip.
\gdef\arghdROWskip#1{%
%  Deletes the next balanced, undelimited argument from a
%                 token list.
% \out{---> Entering arghdROWskip <---}
% \out{In arghdROWskip:  deleted arg is [#1]}%
   \let\next\ROWcount \next%
}%  End of macro \arghdROWskip.
\gdef\argCOLskip#1{%
%  Deletes the next balanced, undelimited argument from a
%                 token list.
% \out{---> Entering argCOLskip <---}
% \out{In argCOLskip:  deleted arg is [#1]}%
   \let\next\COLcount \next%
}%  End of macro \argskip.
}%  End of active &'s.
}%  End of active |'s.
\def\spancount#1{%\out{spancount--->\meaning#1}
   \nspan=#1\multiply\nspan by 2\advance\nspan by -1%
   \global\advance \countREGISTER by \nspan
%  \out{number spancount--->\the\nspan; \the\countREGISTER}
   \let\next\COLcount \next}%
\def\dvr#1{\relax}%
% \omit\hfil%
% \parindent=0pt\hsize=1.1in\valign{%
% \vfil#\vfil&\vfil#\vfil\cr\hfil\hbox{\ Added to\ }\hfil&%
% \hfil\hbox{\ empty events\ }\hfil\cr}\hfil%
\def\header#1{%
\dvr{1}{\let\cr=\@mpersand%
\hdtks={#1}%
%\out{In header:  hdtks=[\the\hdtks]}%
\counthdROWS\hdtks\into\hdrows%
\advance\hdrows by 1%
\ifnum\hdrows=0 \hdrows=1 \fi%
%\out{In header:  Nhdrows=[\the\hdrows]}%
\dvr{5}\makehdPREAMBLE{\the\hdrows}%
%\out{In header:  headerpreamble=[\headerpreamble]}%
\dvr{6}\getHDdimen{#1}%
%\out{In header:  hdsize=[\the\hdsize]}%
%\striplastCR{#1}%
{\parindent=0pt\hsize=\hdsize{\let\ifmath0%
\xdef\next{\valign{\headerpreamble #1\crnorm}}}\dvr{7}\next\dvr{8}%
}%
}\dvr{2}}%  End of macro \header.
\def\makehdPREAMBLE#1{%This macro generates the necessary preamble for a
\dvr{3}%
%                      ruled table with \ncols primary columns.
%                      (Primary columns means the number of columns NOT
%                       counting those used for vertical rules.
\hdrows=#1%  Get the number of columns desired.
{%  Start local parameter definitions.
\let\headerARGS=0%
%  This is the key to the whole thing; it prevents \ARGS
\let\cr=\crnorm%
%                from being expanded in the followin \edef's.
\edef\xtp{\vfil\hfil\hbox{\headerARGS}\hfil\vfil}%
\advance\hdrows by -1%  One row has been generated; decrement the
%                         counter.
\loop%  Append as many further rows as needed to the preamble.
\ifnum\hdrows>0%
\advance\hdrows by -1%
\edef\xtp{\xtp&\vfil\hfil\hbox{\headerARGS}\hfil\vfil}%
\repeat%
\xdef\headerpreamble{\xtp\crcr}%
}%  End of local parameters.
\dvr{4}}%  End of \makehdPREAMBLE.
\def\getHDdimen#1{%
%\out{In getHDdimen:  Arg 1=[#1]}%
\hdsize=0pt%
\getsize#1\cr\end\cr%
}%  End of macro getHDdimen.
\def\getsize#1\cr{%
%\out{In getsize:  Arg 1=[#1]}%
%  Here we have to check arg#1 and see if the first token in #1 is an
%    \end; if so, we stop, else we check the width of arg#1.
%  We recall that each arg#1 will be terminated with a \cr token.
\endsizefalse\savetks={#1}%
%\out{In getsize:  the savetks = [\the\savetks]}%
\expandafter\lookend\the\savetks\cr%
%\out{In getsize:  ifendsize = [\meaning\ifendsize]}%
\relax \ifendsize \let\next\relax \else%
\setbox\hdbox=\hbox{#1}\newhdsize=1.0\wd\hdbox%
\ifdim\newhdsize>\hdsize \hdsize=\newhdsize \fi%
%\out{In getsize:  hdsize=[\the\hdsize]}%
%\out{In getsize:  newhdsize=[\the\newhdsize]}%
\let\next\getsize \fi%
\next%
}%
\def\lookend{\afterassignment\sublookend\let\looknext= }%
\def\sublookend{\relax%
%\out{In sublookend:  looknext = [\looknext]}%
\ifx\looknext\cr %
%\out{In sublooknext:  looknext=cr}%
\let\looknext\relax \else %
%\out{In sublooknext:  looknext/=cr}%
   \relax
   \ifx\looknext\end \global\endsizetrue \fi%
   \let\looknext=\lookend%
    \fi \looknext%
}%
%
%  Allow the user to make his own names for crthick, etc.
%
\def\tablelet#1{%
   \tableLETtokens=\expandafter{\the\tableLETtokens #1}%
}%
\catcode`\@=12%  Change @'s back to their normal category code.

\preprintno{NI 92016, OUTP 92-54S, UCSBTH-92-50}

\title
Universal Amplitude Combinations for Self-Avoiding Walks, Polygons and Trails
\author
John L. Cardy$^*$
\affil
Isaac Newton Institute for Mathematical Sciences
20 Clarkson Road
Cambridge CB3 0EH, UK
\author Anthony J. Guttmann$^{\dag}$
\affil
Department of Physics
Theoretical Physics
University of Oxford
1 Keble Road
Oxford OX1 3NP, UK
\abstract
We give exact relations for a number of amplitude
combinations that occur in the study of self-avoiding walks, polygons
and lattice trails. In particular, we elucidate the lattice-dependent
factors which occur in those combinations which are otherwise
universal, show how these are modified
for oriented lattices, and give new results for amplitude
ratios involving even moments of the area of polygons.
We also survey numerical results for a wide range of amplitudes on a number
of oriented and regular lattices, and provide some new ones.

\vfill
\singlespace
\noindent$^*$ Permanent address: Dept. of Physics, University of California,
Santa Barbara CA 93106, USA

\noindent$^{\dag}$ Permanent address: Dept. of Mathematics, University of
Melbourne, Parkville, Vic 3052, Australia
\eject
\body

\head{1. Introduction.}

In this paper we consider a number of amplitude combinations that
are universal, up to explicit lattice-dependent factors,
in the context of the $N\to0$
limit of the $O(N)$ model, which describes self-avoiding walks
and polygons. We
correct and extend a table of such values given previously by one of
us\refto{TG1}, and correct and generalise an exact expression
for a particular amplitude combination given by the
other\refto{JC}. We also point out an apparent error in another
amplitude combination given previously\refto{PR}.
We make clear the role that lattice-dependent factors (such as the
number of sites per unit area) play in some of these otherwise
universal combinations. In particular, we point out for the first
time a distinction between oriented and unoriented lattices.
In order to confirm these results, we have
generated short series for the radius of gyration of self-avoiding
polygons on the triangular and honeycomb lattices. We also give
arguments for the universality and lattice independence of amplitude
ratios involving even moments of the area of self-avoiding polygons.

The functions we are considering are:
(i) the chain generating function for SAWs, $C(x) = \Sigma c_n x^n$;
(ii) the corresponding polygon generating function, $P(x) = \Sigma p_n x^n$;
(iii) the generating function for lattice trails, $T(x) = \Sigma t_n x^n$;
(iv) the generating function for dumb-bell graphs $\Delta(x) = \Sigma d_n
x^n$;
(v) the mean-square end-to-end distance of $n$-step self-avoiding walks
(SAWs) $\langle
R^2_e\rangle_n$;
(vi) the mean-square radius of gyration of $n$-step polygons $\langle
R^2 \rangle_n$;
(vii) the mean-square radius of gyration of $n$-step SAWs $\langle
R^2_g \rangle_n$;
(viii) moments of the area of polygons of perimeter $n$, $\langle
a^p\rangle_n$;
and (ix) the mean-square distance of a monomer from the origin of $n$-step SAWs
$\langle
R^2_m \rangle_n$.
In the above generating functions, $c_n, p_n, t_n$ and $d_n$ denote the
total number of $n$-step SAWs, polygons, trails and dumb-bells. For SAWs and
trails, an origin is chosen on an infinite lattice, and all distinct SAWs
and trails are enumerated. For dumb-bells and polygons, we adopt the normal
convention in which the total number {\it per lattice site} is given. As we
discuss below, this is different from the number of distinct unrooted
polygons (up to translations) on certain lattices, such as the honeycomb
lattice. This distinction is one source of error in \ref{TG1}.

In terms of the quantities defined above, we denote the relevant amplitudes
as follows:
$$
\eqalignno{
c_n&= A\mu^n n^{\gamma-1}[1 + o(1)]\cr
p_n&= B\mu^n n^{\alpha-3}[1 + o(1)]\cr
t_n&= H\mu^n n^{\gamma-1}[1 + o(1)]\cr
d_n&= J\mu^n n^{\gamma-1}[1 + o(1)]\quad;\cr}
$$
(where the growth exponent $\mu$ for trails applies only to lattices
of coordination number three\refto{GW}), and
$$
\eqalignno{
\langle R^2_e \rangle_n&= Cn^{2\nu}[1 + o(1)]\cr
\langle R^2 \rangle_n&= Dn^{2\nu}[1 + o(1)]&(F)\cr
\langle  a^p\rangle_n&= E^{(p)}n^{2p\nu}[1 + o(1)]\cr
\langle R^2_g \rangle_n&= Fn^{2\nu}[1 + o(1)]\cr
\langle R^2_m \rangle_n&= Gn^{2\nu}[1 + o(1)]\quad;\cr}
$$
where $\gamma = 43/32$, $\alpha = {1 \over 2}$, $\nu = {3 \over 4}$ and $\mu =
x^{-1}_c$, the reciprocal of the critical point.
Amplitudes $B$, $D$ and $E^{(p)}$,
which relate to polygons, will be zero for odd-order terms on loose packed
lattices, and will be non-zero only for every fourth term on certain
lattices such as the L-lattice (an oriented square lattice in which every
step must be followed by a step perpendicular to the preceding step)
and the Manhattan lattice. We should remind the reader that the above
asymptotic forms, and the exact values of the critical exponents, are
assumptions in the sense that they have not been proven rigorously.
Nevertheless they all follow from the central assumption that these problems
have
a continuum limit which corresponds to a particular exactly soluble
field theory, and it is this correspondence on which shall base our
analytic results. The appearance of the same growth exponent $\mu$
in each of the above has been rigorously demonstrated, however.

In the next section we generalise and correct the first amplitude relation,
which gives the value of $BD$. We subsequently discuss the other known
relations, and correct the relation for the combination $BC$. We then use
 these to provide amplitude estimates for some
cases in which they have not been directly estimated.

\head{2. The combination $BD$.}

Rather than giving the modifications to the
argument in \ref{JC} (which in fact contains several errors of factors
of 2), it is clearer to repeat the whole argument for a general lattice.
The universality of the combination $BD$ follows from
the integral form\refto{JCC} of Zamolodchikov's $c$-theorem\refto{ZAM},
which reads
$$
c=3\pi t^2\nu^{-2}\int r^2\langle E(r)E(0)\rangle_cd^2\!r
\quad,
\eqno(1)
$$
where $E(r)$ is the energy density, which enters the
Hamiltonian in the form $t\int E(r)d^2\!r$, and $\nu$ is the usual
critical exponent governing the divergence of the correlation
length. We wish to apply this
to the $O(N)$ model whose lattice Hamiltonian is
${\cal H}=-x\sum_{\rm links}E_{\rm lat}(r)$ where $r$ now labels
links, and $E_{\rm lat}(r)=\vec s(r_<)\cdot\vec s(r_>)$. Here $\vec s(r_<)$
and $\vec s(r_>)$ are $O(N)$ spins located at the sites $r_<$ and $r_>$
at the ends of the link at $r$, ordered in some standard fashion.
In order to apply \(1) correctly, we must relate the continuum energy
density $E(r)$ to its lattice counterpart.
This is done by equating the continuum and lattice Hamiltonians,
so that
$$
t\int E(r)d^2\!r\longrightarrow(x_c-x)\sum_rE_{\rm lat}(r)  \quad.\eqno(2)
$$
It is convenient to work temporarily on a very large but finite lattice
of total area ${\cal A}$, and to rewrite \(1), using translational invariance,
as
$$
c=3\pi t^2\nu^{-2}{\cal A}^{-1}\int\int(r-r')^2\langle E(r)E(r')\rangle_c
\,d^2\!rd^2\!r'
\quad.
$$
It is then apparent from \(2) that we may simply replace this by its
lattice version
$$
c=3\pi(x_c-x)^2\nu^{-2}
{\cal A}^{-1}\sum_{r,r'}(r-r')^2\langle E_{\rm lat}(r)E_{\rm lat}
(r')\rangle_c
\quad.
\eqno(4)
$$
This should be valid in the scaling limit as $x\to x_c$, when the integral
will be dominated by values of $|r-r'|$ on the scale of the correlation
length $\xi$, which is much larger than the lattice spacing, so that the
continuum approximation becomes arbitrarily accurate.

The next step is to evaluate the right hand side of \(4) in the limit $N\to0$
as a sum over pairs of mutually self-avoiding walks connecting $(r_<,r_>)$
and $(r'_<,r'_>)$. These pairs of walks are then identified with self-avoiding
polygons where the links at $r$ and $r'$ are marked. The contribution of
a given polygon of length $n$, whose links are at $(r_1,r_2,\ldots,r_n)$,
to the sum in \(4) is then
$$
Nx^{n-2}{\sum_{k,l}}'(r_k-r_l)^2
\quad,
\eqno(5)
$$
where the prime on the sum indicates that adjacent links are to be excluded.
This excludes contributions where, for example, $r_>$ coincides with $r'_<$.
Such contributions are certainly included in the right hand side of \(4),
but they are more complicated to evaluate. Fortunately they are expected
to be negligible in the scaling limit. It is worth checking the normalisation
in \(5) for small polygons. For example, for $n=4$ on a square lattice, the
sum on the right hand side of \(4) gives $2N{\cal N}_bx^2$, where ${\cal N}_b$
is the
total number of links, while \(5) gives $4Nx^2$ for each elementary square,
of which there are ${\cal N}_b/2$. It is straightforward to check this also for
larger polygons.
Now, if the sum in \(5) were unrestricted, it would be equal to
$2n^2R^2_b$, where $R^2_b$ is the link-weighted squared radius of gyration of
the polygon (which differs from the corresponding site-weighted quantity
by terms of $O(1)$ as $n\to\infty$.) However, the effect of the restriction
is only at the level of subleading terms
down by one power of $n$,
and is negligible in the scaling limit.
If now we denote the mean square radius of gyration over all polygons
of perimeter $n$ by $\langle R^2\rangle_n$, and the total number of such
polygons by ${\cal N}_sp_n$, where ${\cal N}_s$ is the total number of sites on
our large
lattice, then, in the limits $N\to0$ and $x\to x_c$,
$$
c(N)\sim 6\pi N(x_c-x)^2\nu^{-2}\left({{\cal N}_s\over{\cal A}}\right)
\sum_nn^2p_n\langle R^2\rangle_nx^{n-2}
\quad.
\eqno(6)
$$
This implies that the sum on the right hand side has a singularity
of the form $(x_c-x)^{-2}$. Since each term in these series is
non-negative, this singularity on the positive real axis gives
the radius of convergence, and, for
close-packed lattices, we assume that there
 are no other singularities
on $|x|=x_c$. Thus this singularity
solely dictates the behaviour of the terms as $n\to\infty$.
For loose-packed and oriented lattices, however,
$p_n$ is non-zero only if $n$ is
divisible by an integer $\sigma$. In this case the generating function
is invariant under rotations $x\to xe^{2\pi i/\sigma}$, so that there
must exist singularities of equal strengths at $x=x_ce^{2\pi ir/\sigma}$
where $r=0,1,\ldots,\sigma-1$. Thus, the large $n$ behaviour of $p_n$,
for $n$ divisible by $\sigma$, is greater by a factor $\sigma$ than that
expected on the basis of the singularity described by \(6).

We conclude that
$$
\lim_{n\to\infty}np_n\langle R^2\rangle_nx_c^n=
\sigma a_0\,{c'(0)\nu^2\over6\pi}
\quad,
$$
where we have introduced the area per site $a_0={\cal A}/
{\cal N}_s$.
{}From the known values\refto{BCN} $c=1-6/m(m+1)$ and $\nu=(m+1)/4$,
where $N=2\cos(\pi/m)$, for the $O(N)$ model, the
second numerical factor is $5/32\pi^2$,
so that, in terms of the amplitudes defined previously, we have
$$
BD={5\over 32\pi^2}\,\sigma a_0
\quad.
\eqno(z)
$$
For the square lattice, $\sigma=2$ and the area per site (measured in
units of square lattice spacings) is unity. This then gives the result
$5/16\pi^2$ found previously\refto{JC}, which agrees with results
from enumerations given in \ref{PRUD}, as well as later data summarised
in the final section.

For a general periodic lattice, we may write the factor $a_0=
{\cal A}/{\cal N}_s$ as
$v/\kappa$, where $v$ is the area of the unit cell, and $\kappa$ counts
the the number of sites per unit cell.
In the above, we have defined $p_n$ in the conventional way
as the number of $n$-step polygons
per site. If instead, as in \ref{PR},
it is defined as the number of {\it distinct}
$n$-step unrooted polygons (where two such polygons are regarded as
being equivalent if they are related by translation by a lattice vector)
then we may associate with each such polygon a unique site. This may be
done, for example, by ordering the sites in a systematic manner, and
associating with a given polygon the lowest site according to this ordering.
In this way of counting,
${\cal N}_s$ should be replaced by the total number of sites
which are equivalent to each other by lattice translations. These lattice
translations divide the sites into equivalence classes, and there is just
one representative of each equivalence class corresponding to each unit cell.
Thus, with this definition of $p_n$, the factor $\kappa$
does not enter the formula for $BD$.
This is most
easily seen by considering polygons on the honeycomb lattice. The
first term in the generating function is $p_6 = {1 \over 2}$ if the
normalisation is per site, and then $\kappa = 2$. Alternatively,
$p_6 = 1$ gives the number of $6$-step unrooted polygons.

This result may easily be generalised to polygons on oriented lattices.
In this case each link $r$ has a specified orientation, say from $r_<$ to
$r_>$. Oriented walks on such a lattice are described by the $N\to0$ limit
of a {\it complex} $O(N)$ spin model,
where now the link energy is
$E^{\rm lat}(r)=\vec s(r_>)^*\cdot\vec s(r_<)$. Proceeding as before, the
sum on the right hand side of \(4) is now equal to
$$
N\sum_nx^{n-2}\sum_{{\rm all}\ n-{\rm step}\atop{\rm oriented\ polygons}}
{\sum_{k,l}}'
(r_k-r_l)^2
\quad,
$$
so that the appropriate lattice-dependent factor is the total number of
$n$-step oriented polygons divided by the total
area. If we define $p_n$ as the number of $n$-step oriented
polygons per site, then this factor is equal to $p_n/a_0$, as before.
However, the complex $O(N)$ model is equivalent to a real $O(2N)$
model, and therefore the appropriate value of the central charge
on the left hand side of \(6) is $c(2N)$. This leads to an extra
factor of 2 on differentiating with respect to $N$ in the $N\to0$
limit, so the final result for oriented lattices is
$BD/\sigma a_0=5/16\pi^2$, as compared with \(z) for the unoriented
case. If we define $p_n$ as the number of distinct unrooted
oriented polygons, as in \ref{PR}, then $a_0$ should be replaced
by $v$, the unit cell size, in the above formula.

The results for unoriented and oriented lattices may be unified
by supposing that in the unoriented case each link may be oriented
in either direction,
and by defining $p_n$ in both cases
in terms of the number of oriented polygons.
The values of $p_n$ for unoriented lattices will then be
doubled, since each polygon on such a lattice has precisely two
orientations. The result above will then apply to both types of
lattice, and, presumably, to partially oriented lattices also.
It makes physical sense, because
if one looks at polygons with a fixed $R^2$, then the number
of oriented polygons on an oriented lattice should be asymptotically
the same as
the number of such polygons on an unoriented, or partially
oriented, lattice.

\head{3. Polygon area amplitudes.}

Next we consider the universality of amplitudes involving the
area of polygons.
Our approach is as follows:
let $\overrightarrow{dl}$ be an infinitesimal element of the curve
formed by the polygon. Then, by Stokes' theorem, the area within the
polygon is proportional to $\oint\vec r\times\overrightarrow{dl}$
We can write this equivalently as $\oint(\vec r\times\vec J)dl$,
where $\vec J$ is a unit current flowing through the
links of the polygon.
Thus, moments of the area distribution are related
to integrals over
correlation functions of $\vec J$. Such a current will be conserved
in the ensemble of polygons.
In the continuum
field theory, it turns out that it is possible to identify the
corresponding current. Conserved currents in field theory enjoy
the special property that their correlation functions scale according
to naive dimensional counting arguments. Moreover, the normalization
of this particular
current is fixed by considering its behavior in the larger
ensemble including self-avoiding walks with ends. By definition, each end
will be a unit source or sink for this current. Thus the functional
form, and overall normalization, of these correlation functions are
in principle completely determined in the field theory.

We now put some flesh on these remarks.
Consider a given oriented polygon, and let $\vec J^{\,\rm lat}=(J^{\rm lat}_1,
J^{\rm lat}_2)$
be a unit current flowing along the links, each of length $b$ and
labelled by $\vec r$,
in the direction of their orientation.
Then the sum
$$
a=\ffrac12b\sum_{\vec r}\epsilon_{ij}r_iJ^{\rm lat}_j
\quad,
$$
where $\epsilon_{ij}$ is the totally antisymmetric symbol,
gives the {\it signed} area of the polygon (that is, the area is positive
for anticlockwise orientation, negative for clockwise orientation.)
The mean signed area over all polygons with $n$ steps is of course zero,
since opposite orientations of a given polymer give an equal and opposite
contribution. However, the mean square area, and, indeed, all the higher
even moments,
are non-zero and the same as for unoriented polygons.
The squared area of a given polygon is
$$
a^2=\ffrac14b^2\sum_{r,r'}\bigg(r_ir_i'J^{\rm lat}_j(r)
J^{\rm lat}_j(r')-r_ir_j'J^{\rm lat}_i(r)J^{\rm lat}_j(r')\bigg)
\quad.
$$
Using $\sum_rJ^{\rm lat}_i(r)=0$ and $\sum_rr_iJ^{\rm lat}_i(r)=0$, the
right hand side
may be rewritten as $-\frac18b^2\sum_{r,r'}(r-r')^2
J^{\rm lat}_j(r)J^{\rm lat}_j(r')$.

Oriented self-avoiding polygons and walks are described by
a complex $O(N)$ spin model in the limit $N\to0$. The lattice degrees of
freedom are complex $O(N)$ spins, and the Hamiltonian is
${\cal H}=-x\sum_r\vec s(r_>)^*\cdot\vec s(r_<)+{\rm c.c.}$.
Within this model, we may identify
$$
\vec J^{\,\rm lat}(r)=
xb^{-1}(\vec r_>-\vec r_<)\bigg(\vec s(r_>)^*\cdot\vec s(r_<)
-\vec s(r_<)^*\cdot\vec s(r_>)\bigg)
\quad.
$$
Thus, if $\langle a^2\rangle_n$ denotes the mean square area of oriented
polygons
of $n$ steps, and, as before, $2{\cal N}_sp_n$ is the number of such polygons
(the factor of 2 is correct on an unoriented lattice, if $p_n$ counts the
number of unoriented polygons per site),
$$
2{\cal N}_s\sum_np_n\langle a^2\rangle_nx^n=
-\ffrac18b^2\sum_{r,r'}(r-r')^2
\langle \vec J^{\,\rm lat}(r)\cdot\vec J^{\,\rm lat}(r')\rangle
\quad.
\eqno(c)
$$

In the continuum limit, $\vec J^{\,\rm lat}$ is replaced by a conserved
current $\vec J$, which is the Noether current for the $U(1)$ transformations
in the complex $O(N)$ field theory which commute with the $O(N)$
rotations\refto{MILLER}.
The connection between them is simply
$$
\sum_r\vec J^{\,\rm lat}(r)\longrightarrow\int\vec J(r)d^2\!r
\quad,
$$
whereby the double sum over $r$ and $r'$ in \(c) becomes a double integral,
which, by translational invariance, equals
$$
-\ffrac18{\cal A}b^2\int r^2\langle\vec J(r)\cdot\vec J(0)\rangle d^2\!r
\quad.
\eqno(e)
$$

Now the main point is that the normalisation of $\vec J$ in the continuum limit
is fixed by the requirement that the unit source for this current
is the free
end of an oriented SAW\refto{MILLER}. This means that if $\phi(r)$ is the
continuum
field corresponding to $s(r)$ (suppressing $O(N)$ indices),
representing the magnetisation density of the $O(N)$ model,
then
$$
\eqalign{
\int\vec J(r')\cdot\overrightarrow{dS}_{r'}\,\phi(r)&=\phi(r)\quad,\cr
\int\vec J(r')\cdot\overrightarrow{dS}_{r'}\,\phi(r)^*&=-\phi(r)^*\quad,\cr}
\eqno(f)
$$
where the integral is over the boundary of a small neighbourhood of $r$,
and $\overrightarrow{dS}_{r'}$ is the outward pointing normal.
\(f) is supposed to make sense when inserted into correlation functions
with other operators. It specifies\refto{MILLER} the normalisation of the pole
term in
the short-distance expansion of $\vec J$ with $\phi$ and $\phi^*$.

We see from \(f) that $\vec J$ has unit scaling dimension
(since $\vec J$ is a conserved current which generates a symmetry of the
theory, it has no anomalous dimension.)
The fact that the normalisation of $\vec J$ is fixed, and therefore
does not contain any metric factors, implies that its two-point function
has the universal form
$$
\langle\vec J(r)\cdot\vec J(0)\rangle=\xi^{-2}f(r/\xi)
\quad,
\eqno(g)
$$
where $\xi$ is the correlation length. Therefore the integral in \(e)
is of the form $U\xi^2\sim U\xi_0^2(1-x/x_c)^{-2\nu}$, where $U$ is a
universal number, in principle
calculable from the underlying continuum field theory,
and $\xi_0$ is the non-universal metric factor in the correlation length.

It then follows from \(c,g) that
$$
p_n\langle a^2\rangle_n\sim
\sigma a_0 U'\xi_0^2\,n^{2\nu-1}x_c^{-n}
\quad,
$$
where $U'$ is universal. Now we know that $\langle R^2\rangle_n$
is related to the ratio of the second to the zeroth moments of the
energy-energy correlation function in the $O(N)$ model, which is of the form
$U''\xi^2$, where again $U''$ is universal. Therefore the amplitude $D$
defined in \(F) has the form $U''\xi_0^2$.
Also $p_n\sim Bn^{-2\nu-1}x_c^{-n}$.
Combining these, we find
$$
{\langle a^2\rangle_n\over\langle R^2\rangle_n^2}\sim
b^2U'''\,{\sigma a_0\over BD}
\quad.
$$
But we argued above that the combination $BD/\sigma a_0$ should
be universal and lattice independent. This establishes the universality
and lattice independence of the ratio on the left hand side, provided
the area is measured in square lattice spacings.

If we now consider the higher even moments of the area, the argument
generalises straightforwardly.
Schematically,
$$
{\cal N}_s\sum_np_n\langle a^{p}\rangle_nx^n\sim b^p\int\ldots\int
 r_1\ldots r_p
\langle J(r_1)\ldots J(r_p)\rangle d^2\!r_1\ldots d^2\!r_p
\quad,
$$
where all indices have been suppressed. The integral on the
the right hand side is equal to ${\cal A}U^{(p)}\xi^{2p-2}$, where $U^{(p)}$
is universal. Following the argument through, we then discover that
$\langle a^{p}\rangle_n/\langle R^2\rangle_n^{p}$ should be universal,
with no lattice dependent factors.

However, it appears difficult to extend this result to odd moments
of the unsigned area. This is because these quantities appear not
to have any simple expression in terms of correlation functions in the
continuum field theory.
Nevertheless, in \ref{CF} it was argued that the corresponding ratio involving
the
first moment of the area
$\langle a^1\rangle_n/\langle R^2\rangle_n = E^{(1)}/D$ should be universal,
with no lattice dependent factors.
This is to be expected on physical grounds\refto{F},
since if lattice polygons are
regarded as a model for 2-dimensional vesicles, the area couples to the
pressure $p$. Since $p$ and $\langle R^2\rangle$ are macroscopically
measurable quantities, one might expect a universal relationship
between them independent of the particular lattice model.
Nevertheless, a theoretical demonstration of this is so far lacking.
In \ref{DUP}, an argument was made for the finiteness of ratio
$E^{(1)}/D$, but this does not prove its universality.
Note that\refto{F} the fact that moments of the area satisfy the
inequalities
$$
\langle a^1\rangle\leq\langle a^2\rangle^{\frac12}\leq\langle a^3\rangle
^{\frac13}\leq\ldots
\quad,
$$
together with our results for the even moments, does imply that the
higher
odd moments $\langle a^{p}\rangle$ do scale as expected for $p\geq3$,
but it does not imply that their ratios are universal or lattice independent.
As pointed out\refto{F}, it is essential that
area be measured in Euclidean units such that the lattice spacing is the same
in
calculating the radius of gyration and the mean area, nd its moments.

\head{4. Numerical results.}

Another universal amplitude ratio, first given in \ref{CS} and corrected
in \ref{CPS} is

$$(2 + {y_t \over y_h}){F \over C} -{2G \over C} + {1 \over 2} = 0
\quad.$$
The individual quotients $F/C$ and $G/C$ are also universal
\refto{CS}.
The best estimates of these quantities is given in \ref{CPS}, as
$F/C=0.14029\pm0.00012$ and $G/C=0.43962\pm0.00033$,
obtained by Monte Carlo calculations.

The quantity $BC/\sigma a_0$ has been shown in \ref{PR}
to be universal. These authors expressed their result in terms of
$BC/v\sigma$, but they used a definition of $p_n$ as the number of
distinct unrooted polygons. As discussed earlier, these results
are completely equivalent. It is possible to show directly
from the scaling forms of the relevant correlation functions in
the $O(N)$ model that $C/D$ is universal. This also follows from
\ref{CS} and \ref{PHA}, where it was argued
that $F/C$ and $F/D$, respectively, are universal.
Hence, the universality of $BC/\sigma a_0$ follows from our result
for $BD/\sigma a_0$.

The value of $C/D$ should also be the same for walks on oriented
lattices. Since we have argued that in this case $BD/\sigma a_0$
gains an additional factor of two, this should carry over to
$BC/\sigma a_0$.

The amplitude for dumb-bells, $J$, is related to the amplitude for SAWs $A$ by
$$J = A(1 - 2\tau/\mu + \tau^2/\mu^2)/8\quad, \eqno(l)$$
as shown in \ref{TG2}, where
$\tau = q-1$ and $q$ is the lattice coordination number. It
was also shown there that the amplitude for lattice trails, $H$,
{\it on the honeycomb lattice
only}, is related to the SAW amplitude by
$$H = {4A \over {2+\surd2}}\quad. \eqno(m)$$

Details of the various series and their analysis to estimate critical
amplitudes
have already been discussed in \ref{TG1}. Since then,
extension of the square lattice SAW
series\refto{CG1} and the square lattice trail series\refto{CG2} have allowed
more accurate estimates of these amplitudes. These are given in Table I below.
Corrected amplitudes for the honeycomb lattice are also
given in Table I. To ensure
that the various normalisation factors were correct,
we generated and analysed the
series for the radius of gyratio for triangular lattice and honeycomb lattice
polygons, and estimated the radius of gyration amplitude directly.
These new series are given
in Table II, and the amplitude estimates in Table I.
The quantity $E$(triangular)
was given in \ref{TG1} in units of unit triangles. It is more correctly
given in Table I in units of area, assuming unit lattice spacing,
as also used in the
estimate of other metrically dependent amplitudes. Direct estimates of the
amplitudes $A, B$ and $C$ for the L and Manhattan lattices are also given.
These have been obtained from the series in \ref{EG} and \ref{Ma}. The
polygon counts for Manhattan lattice polygons quoted in \ref{EG} should be
divided by $2$ in order that the correct normalisation per-site be retained.

The square lattice amplitudes are generally the most accurate.
We see that\break
$32\pi^2 BD/5\sigma a_0 = 1.0000$
for the square lattice, $1.020$ for the
honeycomb lattice and $0.997$ for the triangular lattice.
This is perfect agreement,
given the accuracy of the amplitude estimates,
and allows $D$(L, Man.) to be estimated.
Similarly, the invariant $BC/ \sigma a_0 = 0.2168$
(square), $0.2167$ (triangular) and this value is used
to predict $C$(honeycomb). The ratio $F/C = 0.1403$(square), $0.1402$
(triangular), in good agreement with the Monte Carlo estimate
cited above. This then
permits $F$(honeycomb, L, Man.) to be estimated.
We see that this quantity is a factor 2 greater for the oriented
L and Manhattan lattices, as predicted. This factor was not
found by Privman and Redner\refto{PR}, who found that (with their
convention for $p_n$ as the number of distinct unrooted oriented
polygons) the same result for $BC/v\sigma$ as for unoriented lattices, within
their numerical errors. It seems likely
that these authors miscounted by a factor of 2. For example, the number
of distinct oriented 4-step polygons on the L lattice is 2, not 1.

Similarly, $G/C = 0.4397$(square), and
$0.4402$(triangular), again in agreement with the precise
Monte Carlo estimate. The
predicted values of $G$(honeycomb, L, Man) are given in Table I.
The ratio $E/D$ (where $E$ is $E^{(1)}$ in our previous notation) was
found to be $2.515$(square) and $2.529$(triangular),
where we believe the square lattice
estimate to be more precise, as the series from which $E$
was estimated is far longer for
the square lattice. This
value then permits $E$(honeycomb, L, Man.) to be estimated.
The amplitudes $J$ follow from
the amplitudes for $A$ and \(l). The amplitude $H$(honeycomb) follows from
$A$(honeycomb) and \(m). All amplitudes quoted are expected to have an
associated
error confined to the last decimal place quoted. Apart from some minor gaps
for the triangular, Manhattan and L lattice amplitude, Table I now
gives a complete and corrected tabulation of
critical amplitudes.

\noindent{\sl Acknowledgments.}
We would like to thank Prof.~Michael E.~Fisher for valuable
comments and correspondence, and Mr. Andrew Conway for programming assistance
in calculating the radii of gyration.
This work was supported in part
by the Isaac Newton Institute, the UK Science and Engineering Research
Council, the US National Science Foundation Grant PHY 91-16964,
and by the Australian Research Council.

\head{Tables}
\vskip .5in
\singlespace
\noindent{\bf Table I}. Estimates of critical amplitudes
$A\ldots J$ defined in the
text, for the
honeycomb, square, triangular, L and Manhattan
lattices. Quantities in parentheses are estimates
from the amplitude relations discussed in the text. The remainder are
estimates from the corresponding series expansions.
\vskip .5in
\begintable
Amplitude |  Honeycomb |  Square | Triangular | L |Manhattan\crthick
A |     1.145 | 1.1771| 1.186| 1.05|0.89\crnorule
B |     0.6358| 0.5623| 0.2640 | 2.47|2.5\crnorule
C |     (0.889)| 0.771| 0.711 | 0.67| 0.73\crnorule
D |     0.0660| 0.05631| 0.0518 | (0.049)|(0.053)\crnorule
E |     (0.166)| 0.1416| 0.131 | (0.12)|(0.13)\crnorule
F |     (0.125)| 0.1082| 0.0997 | (0.095)|(0.10)\crnorule
G |     (0.389)| 0.339 | 0.313 | (0.30)|(0.32)\crnorule
H |    (1.341)| 1.272 | | |\crnorule
J |     (0.000972)| (0.002768) | (0.006205)| | \cr
$\mu$ |  1.8477591 | 2.6381585| 4.1507951| 1.5657| 1.733\crnorule
$\sigma$ | 2 | 2 | 1 | 4| 4\crnorule
$a_0$| $3\surd3/4$ | 1 | $\surd3/2$ | 1|1\cr
$32\pi^2BD/5\sigma a_0$|1.020|1.0000|0.997| | \crnorule
$BC/\sigma a_0$| |0.2168|0.2167|0.41|0.46\endtable

\vfill\eject
\singlespace
\noindent{\bf Table II}. Number of polygons and the sum of the squares of the
radii
of gyration of n-step polygons on the triangular and honeycomb lattices.
\vskip .5in
\begintable
\multispan3\phantom{xxx}||\multispan3\phantom{xxx}\crnorule
\multispan3 Triangular || \multispan3 Honeycomb \crnorule
\multispan3\phantom{xxx}||\multispan3\phantom{xxx}\cr
$n$|$p_n$|$4n^2 p_n\langle R^2\rangle_n$||$n$|$2p_n$|$8n^2 p_n\langle
R^2\rangle_n$\crthick
3|\hfil2|\hfil24||\hfil6|\hfil1|\hfil 144 \crnorule
4|\hfil3|\hfil96||\hfil8|\hfil0|\hfil0 \crnorule
5|\hfil6|\hfil408||\hfil10|\hfil3|\hfil2460 \crnorule
6|\hfil15|\hfil1872||\hfil12|\hfil2|\hfil3168 \crnorule
7|\hfil42|\hfil8688||\hfil14|\hfil12|\hfil32052 \crnorule
8|\hfil123|\hfil39912||\hfil16|\hfil18|\hfil77976 \crnorule
9|\hfil380|\hfil183264||\hfil18|\hfil65|\hfil420444 \crnorule
10|\hfil1212|\hfil834744||\hfil20|\hfil138|\hfil1310088 \crnorule
11|\hfil3966|\hfil3779064||\hfil22|\hfil432|\hfil5655204 \crnorule
12|\hfil13265|\hfil17013936||\hfil24|\hfil1074|\hfil19291968 \crnorule
13|\hfil45144|\hfil76186320||\hfil26|\hfil3231|\hfil76066992 \crnorule
14|\hfil155955|\hfil339566616||\hfil28|\hfil8718|\hfil268063080 \crnorule
15|\hfil545690|\hfil1507025568||\hfil30|\hfil25999|\hfil1011675420 \crnorule
16|\hfil1930635|\hfil6662739096||\hfil |\hfil |\hfil  \crnorule
17|\hfil6897210|\hfil29355291552||\hfil |\hfil |\hfil \crnorule
18|\hfil24852576|\hfil128932421592||\hfil |\hfil |\hfil \endtable

\vfill\eject

\references

\oneandathirdspace

\refis{JC} J.~L.~Cardy, \journal J. Phys. A, 21, L797, 1988.

\refis{JCC} J.~L.~Cardy, \journal Phys. Rev. Lett., 60, 2709, 1988.

\refis{BCN} Vl.~Dotsenko and V.~A.~Fateev, \journal Nucl. Phys. B, 240,
312, 1984;
H.~W.~Bl\"ote, J.~L.~Cardy and M.~P.~Nightingale,
\journal Phys. Rev. Lett., 56, 742, 1986.

\refis{ZAM} A.~B.~Zamolodchikov, \journal Zh. Eksp. Teor. Fiz., 43,
565, 1986;[\journal JETP Lett., 43, 730, 1986.]

\refis{CF} C.~J.~Camacho and M.~E.~Fisher, \journal Phys. Rev. Letts., 65, 9,
1990.

\refis{TG1} I.~G.~Enting and A.~J.~Guttmann, \journal J. Phys. A, 25, 2791,
1992.

\refis{PR} V.~Privman and S.~Redner, \journal J. Phys. A, 18, L781, 1985.

\refis{PRUD} V.~Privman and J.~Rudnick, \journal J. Phys. A, 18, L789, 1985.

\refis{PHA} V.~Privman, P.~C.~Hohenberg and A.~Aharony in {\sl Phase
Transitions and
Critical Phenomena} 14, 1, 1991 (eds. C.~Domb and J.~Lebowitz, Academic,
London).

\refis{CPS} S.~Caracciolo, A.~Pelisetto and A.~D.~Sokal, \journal J. Phys. A,
23, L969,
1990.

\refis{CS} J.~L.~Cardy, and H.~S.~Saleur, \journal J. Phys. A, 22, L601, 1989.

\refis{TG2} A.~J.~Guttmann, \journal J. Phys. A, 18, 567, 1985.

\refis{F} M.~E.~Fisher, private communication.

\refis{CG1} A.~R.~Conway, I.~G.~Enting and A.~J.~Guttmann {\sl Algebraic
Techniques for
enumerating SAWs on the square lattice}, {\sl J. Phys. A}, (to appear).

\refis{CG2} A.~R.~Conway and A.~J.~Guttmann {\it Enumeration of Self-avoiding
Trails on
a square lattice}, {\sl J. Phys. A}, (to appear).

\refis{MILLER} J.~Miller, \journal J. Stat. Phys., 63, 89, 1991.

\refis{EG} I.~G.~Enting and A.~J.~Guttmann, \journal J. Phys. A, 18, 1007,
1985.

\refis{Ma} A.~Malakis, \journal J. Phys. A, 8, 1885, 1973.

\refis{DUP} B.~Duplantier, \journal Phys. Rev. Lett., 64, 493, 1990.

\refis{GW} A.~J.~Guttmann and S.~G.~Whittington, \journal
J. Phys. A, 11, 721, 1978.

\endreferences
\endit